\documentclass[12pt]{article} 

\expandafter\let\csname equation*\endcsname\relax
\expandafter\let\csname endequation*\endcsname\relax
\usepackage{amsmath,amssymb,mathtools,amsfonts,latexsym,bm}
\usepackage{graphicx}
\usepackage{float}
\usepackage[usenames,dvipsnames]{color}
\usepackage{times}
\usepackage[english]{babel}
\usepackage[latin1]{inputenc}
\usepackage[T1]{fontenc}
\usepackage{siunitx}

\definecolor{linkcolor}{rgb}{0,0,0.6} 
\usepackage[pdftex,colorlinks=true, pdfstartview=FitV, linkcolor= linkcolor, citecolor= linkcolor, urlcolor= linkcolor, hyperindex=true,hyperfigures=true]{hyperref} 

\renewcommand{\vec}{\bm}
\newcommand{\dd}{{\rm d}}
\newcommand{\mean}[1]{\left< #1 \right>}
\newcommand{\Obs}{\mathcal{O}}

\newcommand{\ie}{{\it i.e}}

\newcommand{\ee}{e}

\newcommand{\del}{\partial}

\newcommand{\Tr}{{\rm Tr}}

\newcommand{\Eqref}[1]{Eq.~\eqref{#1}}

\newcommand{\npf}{B}
\newcommand{\fp}{\alpha}
\newcommand{\dri}{a}
\newcommand{\noi}{\xi}
\newcommand{\dif}{D}
\newcommand{\for}{F}
\newcommand{\mob}{\mu}
\newcommand{\npz}{\tilde{\npf}}
\newcommand{\npd}{\delta \mkern-1mu \npf}
\newcommand{\npdd}{\delta \mkern-1mu \dot{\npf}}
\newcommand{\hh}{h}
\newcommand{\difz}{\tilde{D}}

\newcommand{\obs}{\mathcal{O}}
\newcommand{\xx}{x}
\newcommand{\RF}{\mathcal{R}}
\newcommand{\yy}{\tilde{x}}
\newcommand{\prm}{\lambda}
\newcommand{\sus}{\chi}
\renewcommand{\Pr}{P}
\newcommand{\Act}{\mathcal{A}}

\newcommand{\Acty}{\tilde{\mathcal{A}}}

\newcommand{\driy}{\tilde{\dri}}
\newcommand{\dist}{\rho}
\newcommand{\Lgen}{\mathbb{L}}
\newcommand{\drag}{\gamma}
\newcommand{\Tem}{\Delta T}
\newcommand{\HH}{H}
\newcommand{\Omg}{\Omega}
\newcommand{\GG}{G}

\begin{document}

\title{A general fluctuation-response relation for noise variations and its application to driven hydrodynamic experiments}


\author{Cem Yolcu$^{1}$, Antoine B\'erut$^{2}$, Gianmaria Falasco$^{3,4}$, \\
  Artyom Petrosyan$^{2}$, Sergio Ciliberto$^{2}$, Marco Baiesi$^{1,5,*}$}
\date{\today}

\maketitle

{$^1$ Dipartimento di Fisica e Astronomia ``Galileo Galilei'',
  Universit\`a di Padova, Via Marzolo 8, 35131, Padova, Italy
} \\
{$^2$ Laboratoire de Physique de Ecole Normale Sup\'{e}rieure de Lyon (CNRS UMR5672), 
46 All\'ee d'Italie, 69364 Lyon, France.}\\
{$^3$ Max Planck Institute for Mathematics in the Sciences, Inselstr. 22,
04103 Leipzig, Germany} \\
{$^4$ Institut f\"ur Theoretische Physik, Universit\"at Leipzig, Postfach 100 920, D-04009 Leipzig, Germany} \\
{$^5$ INFN, Sezione di Padova, Via Marzolo 8, 35131, Padova, Italy} \\
{$^*$ email: {\tt baiesi@pd.infn.it}}\\

\begin{abstract}
  The effect of a change of noise amplitudes in overdamped diffusive systems is linked to their unperturbed behavior by means of a nonequilibrium fluctuation-response relation. This formula holds also  for systems with state-independent nontrivial diffusivity matrices, as we show with an application to an experiment of two trapped and hydrodynamically coupled colloids, one of which is subject to an external  random forcing that mimics an effective temperature. The nonequilibrium susceptibility of the energy to a variation  of this driving is an example of our formulation, which improves an earlier version, as it does not depend on the time-discretization of the stochastic dynamics. This scheme holds for generic systems with additive noise and can be easily implemented numerically, thanks to matrix operations.
\end{abstract}




\section*{Introduction}

Several phenomena are often modeled by a continuous stochastic dynamics in
which a noise term randomizes the motion of each degree of freedom.
These noises can have a nontrivial structure. For example, hydrodynamic
interactions between diffusing particles, due to the concerted motion of the
fluid molecules, are represented by a cross-correlation of noise terms
in Langevin equations, with amplitudes proportional to the bath
temperature~\cite{doi1988theory}.

Finding the response of the system to a variation of one of the noise
amplitudes is a task that encounters some mathematical
difficulty. While a variation of deterministic drifts in diffusion
equations may be dealt with by established
tools~\cite{cug94,nak08,che08,spe09,sei10,pro09,ver11a,lip07,bai09,bai09b,alt16}, 
such as the Girsanov theorem~\cite{gir60} and Radon-Nikodym 
derivatives~\cite{kal71,bai09b}, it is not straightforward to compare two dynamics
driven by different noise amplitudes~\cite{bai14,yol16}, due to the
missing absolute continuity between the two path measures.
This might have hampered the definition of a linear response theory to temperature
variations or in general to noise-amplitude variations. 
However, there are recent results in this 
direction~\cite{che09b,bok11,bai14,bra15,pro16,fal16,fal16b,bai16}.
The interest in understanding thermal response in
nonequilibrium conditions is related to the definition of a steady state 
thermodynamics~\cite{oon98,hat01,sek07,sag11,ber13,mae14,kom15,man15},
in which concepts as specific heat~\cite{bok11} are extended to
the realm of dissipative steady states.

To circumvent issues related to the missing Girsanov theorem for
stochastic processes with different noise coefficients,
some attempts to define a thermal linear
response theory needed to resort to a time discretization~\cite{bai14,yol16}.
Recently, with two independent proofs~\cite{fal16,fal16b},
it was proposed a solution avoiding this discretization. Namely, a (continuous time) thermal response formula devoid of singular terms can be obtained either by an explicit regularization procedure based on functional methods \cite{fal16b} or through a coordinate rescaling which turns noise perturbations into mechanical ones \cite{fal16}. 
This formalism was developed for uncorrelated noises and applied to
an experiment of a thermally unbalanced RC circuit~\cite{cil13a,cil13b}, 
determining its nonequilibrium heat capacitance~\cite{bai16}.
However, for example, the scheme described in~\cite{fal16,fal16b} cannot
be applied to hydrodynamically coupled particles.

A recent experiment realized a minimal hydrodynamic system composed of
nearby optically trapped colloids, in which one colloid 
is  shaken by randomly displacing  the optical trap position. 
It has  been shown that this  random displacement plays the role of an effective temperature for the shaken particle, whose motion in some sense is ``hotter'' than that of the other particle~\cite{ber14,ber16}. The result is equivalent to a system in which heat bath as a whole is out of equilibrium.  Not only
does each degree of freedom experience a different temperature, but
also the global structure of the stochastic equations does not meet
the standard form of local detailed balance~\cite{kat83}.  Thus, for
such a system it is not straightforward to define concepts like
entropy production in the environment.  A thermal response in this
context is possible, as shown with a theory including
correlated white noise terms~\cite{yol16}. 
This approach, following the scheme presented in~\cite{bai14}, still
included a time discretization for overcoming the mathematical
difficulties mentioned above, hence it was not yet written in terms
only of sensible mathematical expressions such as (stochastic)
integrals, but also included discrete sums of terms which are
singular in the limit of continuous time.

In this paper we provide the most general thermal response theory for
overdamped diffusive motions with additive correlated white noise,
using a formulation based on path weights~\cite{zinn02}.  We thus merge the
positive aspects of recent approaches in a single, general
framework, which could be used to study how a diffusive process reacts
to a variation of one or many of its noise amplitudes. This formalism
is adopted to analyse the data of the experiment involving
hydrodynamically coupled particles mentioned above, showing how to
apply the scheme in practice. Pragmatically, a matrix formulation
simplifies this last step. In this way, after the previous analysis
of nonequilibrium $RC$ circuits~\cite{bai16}, we continue the application
of a thermal response theory to experimental data. This complements
earlier analysis of experiments focused on the mechanical linear 
response~\cite{gom09,gom11,gom12,boh12}.

Having computed the system's susceptibility
to a variation of the random driving, 
we show that there is a good agreement with another estimate obtained using 
Seifert and Speck's formula~\cite{sei10}.
This is in the class of formulas that focus on the density of 
states or related quantities~\cite{aga72,fal90,spe06,che08,pro09,sei10},
and hence can be general enough to embrace also the thermal response type of problem.
Note that an integrated version of the latter formula~\cite{sei10} 
was recently connected with a statistical
reweighting scheme that reappropriates data obtained from a stationary
experiment as data obtained by an actual perturbation protocol.
Also in this case, one needs to know the steady state distribution.

The following Section introduces the experiment we will analyse.
Dealing first with a real system helps in motivating the need for
the new theory and in exposing the derivation of suitable
fluctuation-response relations (Section~\ref{sec:theory}).
These are the response function to a variation of an element of the inverse
diffusion matrix (\Eqref{eq:R}), and the susceptibility obtained by 
integrating in time a combination of these response functions, see 
\Eqref{eq:suscep}, or \Eqref{eq:sus.matr} for its version in matrix notation.
The susceptibility of the potential energy, either of the system or of the
particle not driven, is derived from the steady state experimental data
in Section~\ref{sec:sus_coll}, which is followed by conclusions and by
further details in appendices.


\section{Experiment} \label{sec:exp}

\begin{figure}[!tb]
\begin{center}
\includegraphics[width=.5\textwidth]{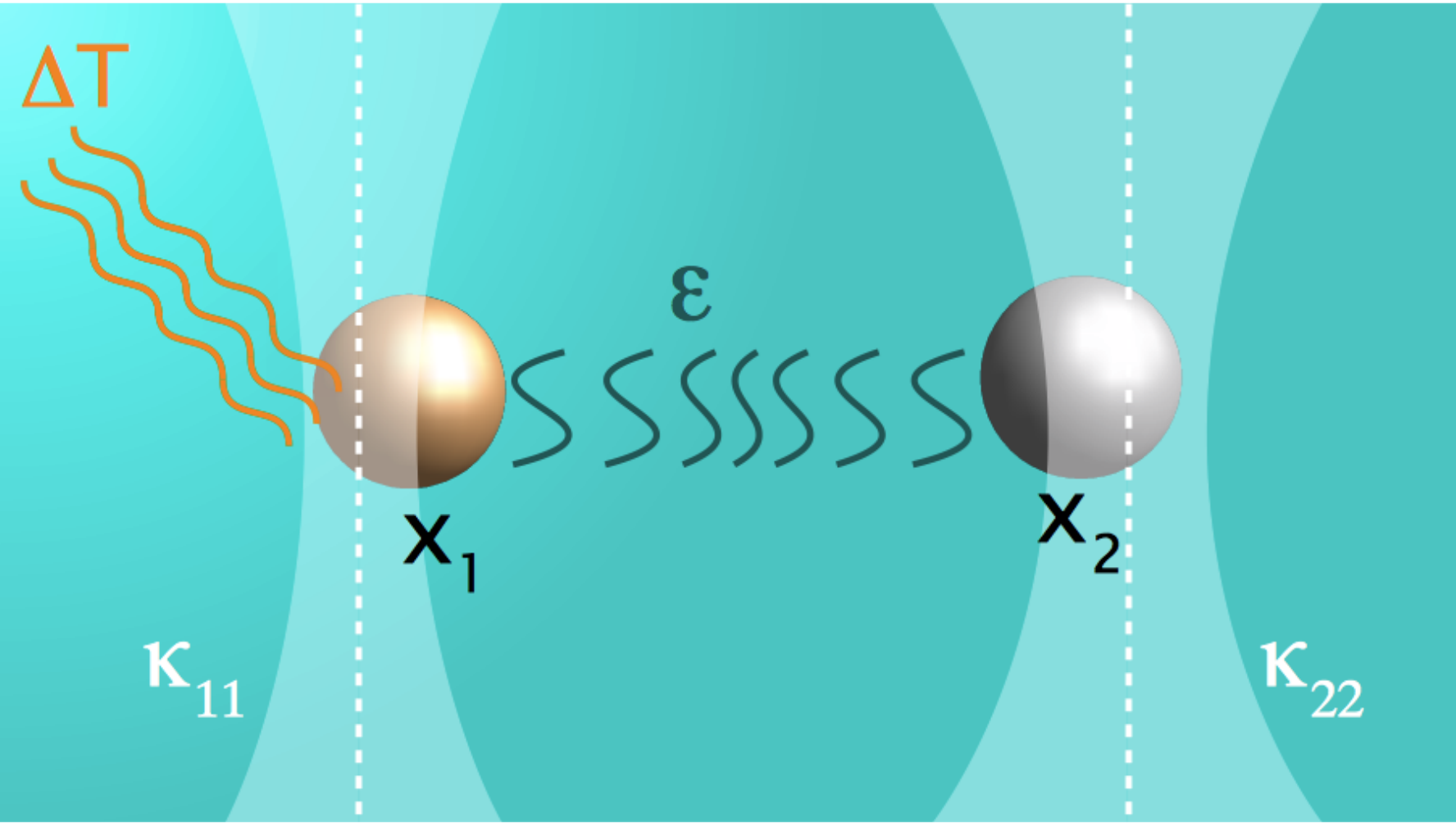}
\end{center}
\caption{(Color online) Sketch of the experiment: two spherical
  colloidal particles in water at temperature $T$
  are confined in independent harmonic potentials
  $(\kappa_{11},\kappa_{22})$ generated by optical traps. The particles
  interact hydrodynamically ($\varepsilon$), and the first particle
  experiences an effective temperature produced by randomly moving the position of the trap. 
  The difference between this effective temperature and the water one ($\Tem$)
  mimics an additional thermal environment setting the system out of
  equilibrium. }
\label{fig:sketch}
\end{figure}

The two particles interaction is studied in  the following
experimental setup. A custom-built vertical optical tweezers with an
oil-immersion objective (HCX PL. APO $63\times$/$0.6$-$1.4$) focuses a
laser beam (wavelength \SI{532}{\nano\meter}) to create a quadratic
potential well where a silica bead (radius $R = \SI{1}{\micro\meter}
\pm 5\%$) can be trapped. The beam goes through an acousto-optic
deflector (AOD) that allows to modify the position of the trap very
rapidly (up to \SI{1}{\mega\hertz}). By switching the trap at
\SI{10}{\kilo\hertz} between two positions we create two independent
traps, which allows us to hold two beads separated by a fixed distance
(a schematic drawing is shown in Figure~\ref{fig:sketch}). The beads
are dispersed in bidistilled water at low concentration to avoid
interactions with multiple other beads. The solution of beads is
contained in a disk-shaped cell (\SI{18}{\milli\meter} in diameter,
$\SI{1}{\milli\meter}$ in depth). The beads are trapped at
$h=\SI{15}{\micro\meter}$ above the bottom surface of the cell. The
position of the beads is tracked by a fast camera with a resolution of
\SI{115}{\nano\meter} per pixel, which after treatment gives the
position with an accuracy better than \SI{5}{\nano\meter}. The
trajectories of the bead are sampled at \SI{800}{\hertz}. The
stiffness of one trap 
($\kappa_{11} = $\SI[separate-uncertainty = true]{3.37(1)}{\pico\newton/\micro\meter} for trap $1$ or
$\kappa_{22} = $\SI[separate-uncertainty = true]{3.33(1)}{\pico\newton/\micro\meter} for trap $2$, each
representing an element of a diagonal stiffness matrix $\kappa$, see
next section) is proportional to the laser intensity\footnote{It can
  be modified by adding neutral density filters or by changing the
  time that the laser spend on each trap.}. The two particles are
trapped on a line (called ``x axis'') and separated by a distance $d$
which is tunable. The experimental results presented here have been
obtained at $d=\SI{3.9}{\micro\meter}$. At this distance  the
Coulombian interaction between the particle surfaces is negligible.

To create an effective temperature on one of the particles (for
example on particle $1$), a Gaussian white noise is sent to the AOD so
that the position of the corresponding trap is moved randomly in the
direction where the particles are aligned. If the amplitude of the
displacement is sufficiently small to stay in the linear regime it
creates a random force on the particle which does not affect the
stiffness of the trap. Here the particles are over-damped and have a
Lorentzian power spectrum with a typical cut-off frequency $f_c$ of
\SI{30}{\hertz}. The added noise is numerically created by a LABVIEW
program: it is sampled at \SI{100}{\kilo\hertz} with a tunable
amplitude $A$ (typically of $\sim \SI{1}{\volt}$) and numerically
low-pass filtered at \SI{1}{\kilo\hertz}. It is then generated by the
analog output of a NI PXIe-6366 card. The conversion factor for the
displacement due to the AOD is \SI{2.8}{\micro\meter/\volt}, and the
typical voltage of the noise after filtration is between $\pm
\SI{0.25}{\volt}$. When the random force is switched on, the bead
quickly reaches a stationary state with an ``effective temperature''
$T_{\rm eff} = T + \Tem$  for the randomly forced degree of
freedom. Here the parameter $\Tem$ quantifies the strength of the
forcing (see \Eqref{eq:dif.2} below) and will be used also as the
tunable strength of the thermal perturbation.  The properties of the
effective temperature $T_{\rm eff}$ have been studied using power
spectra \cite{ber14,Berstat16} and Fluctuations Theorems \cite{ber16}.
This setup allows us to create a wide range of effective temperatures
for one bead, and to look at the interaction between this agitated
bead and another one trapped at equilibrium at a finite distance $d$.


\section{Theory} \label{sec:theory}

Let us consider a system with $N$ degrees of freedom, whose state is a
vector denoted by $x\equiv(x_1,x_2,\ldots,x_N)$.  We write the
equation of motion of our overdamped system in matrix notation as
\begin{align}
  \dot{\xx} = \dri (\xx) + \npf \noi \ , \label{eq:eom}
\end{align}
where the vector $\dri (\xx)$ is the mean drift velocity, and the
$N\times N$ matrix $\npf$ is the noise amplitude, which is assumed to
be independent of the state $\xx$. The differential noise $\noi (t)$
is Gaussian and colorless as well as uncorrelated among different
components, \ie., $\left< \noi_i (t) \noi_j (s) \right> = \delta_{ij}
\delta (t-s)$.  Explicit time dependences of $\xx$, $\noi$, and
possibly $\dri$ and $\npf$, are omitted for simplifying the notation,
except for occasions when emphasis is appropriate. Furthermore, as has
already been hinted at, we will alternate between matrix and index
notations according to whichever is more intelligible in specific
situations.

As an example, the drift vector of a hydrodynamic system is given 
in terms of the deterministic force $\for(\xx)$ and of the 
mobility matrix $\mob$,
\begin{align}
  \dri (\xx) = \mob \for (\xx) \ . \label{eq:dri}
\end{align}
Due to hydrodynamic interactions, the mobility matrix has nonvanishing
off-diagonal entries $\mob_{ij}\ne 0$ that couple different degrees of
freedom $i\ne j$.  In general this would result in a state-dependent
mobility, since the coupling depends on the difference of
coordinates. However, we are concerned with an experiment where a
state-independent mobility is a viable approximation: The hydrodynamic
coupling between distant degrees of freedom is in the form of a series
expansion in the ratio of particle radius to distance
\cite{doi1988theory}. Hence, if the system is maintained such that the
average distance is considerably larger than its fluctuations, it can
be assumed to remain constant at its average value to lowest
order. The optical traps in the considered experimental setup
are ``stiff'' enough to achieve just that. The resulting mobility
matrix, to the aforementioned level of approximation, for
this system with $N=2$ is given as
\begin{align}
  \mob = \frac 1\drag \begin{bmatrix} 1 & \varepsilon \\
    \varepsilon & 1 \end{bmatrix} \ , \label{eq:mob}
\end{align}
where $\varepsilon = 3 R/2d$ is a small hydrodynamic coupling between
spherical colloids of radius $R$ separated by a(n average) distance
$d$ according to the Oseen approximation 
($\varepsilon=0.2766$ in the experiment we consider),
and $\gamma = $\SI{16.8}{\pico\newton \milli \second / \micro\meter}
is the drag coefficient of the colloids
in water.
The distance $d$ is set by the separation of the two harmonic 
optical traps, whose associated potential energy is 
\begin{align}
  U(\xx) = \tfrac 12 \xx^\dagger \kappa \xx \ , \label{eq:U}
\end{align}
when each coordinate is measured with respect to the center of its
corresponding trap, where $\kappa$ is a diagonal matrix of stiffness
constants.

As a result of the form of the mobility matrix, the diffusivity
\begin{align}
  \dif = \tfrac 12 \npf \npf^\dagger \label{eq:dif}
\end{align}
is also endowed with nonvanishing off-diagonal entries that are
state-independent. In fact, in the given experimental setup, the
diffusivity is a sum of two contributions \cite{yol16}
\begin{align}
  \dif = T \mob + \Tem \mob \begin{bmatrix} \drag&0\\0&0 \end{bmatrix}
  \mob = \frac 1\drag \begin{bmatrix} T+\Tem &
    \varepsilon(T+\Tem)\\ \varepsilon (T+\Tem) & T+\varepsilon^2
    \Tem \end{bmatrix} \ , \label{eq:dif.2}
\end{align}
the first of which originates from the solvent at temperature $T$, and
the second from the stochastic forcing on particle $1$, which has a white
power spectrum of magnitude $\drag \Tem$ \cite{ber14}.

\subsection{Thermal perturbation and change of variables}
\label{sec:cov}

Our aim is to investigate the response of the system to a small
variation in its diffusion matrix, which is the most direct analog of
temperature in this multidimensional situation. However, it is better
to begin the treatment in terms of the noise amplitude, which is less
physical, and revert to diffusivity at an appropriate stage.

We begin by perturbing the noise prefactor matrix as,
\begin{align}
  \npf (t) = \npz + \npd (t) \ , \label{eq:pert}
\end{align}
and then follow the technique of Ref.\ \cite{fal16}, which is
unfettered by terms that depend on the time discretization. To this
end, we change variables by defining
\begin{align}
  \xx (t) = \npf (t) \npz^{-1} \yy (t) \ , \label{eq:cov}
\end{align}
which implies
\begin{align}
  \dot{\xx} = \npf \npz^{-1} \dot{\yy} + \npdd
  \npz^{-1} \yy \ .
\end{align}
Plugging this into the equation of motion \eqref{eq:eom}, and
multiplying it by $\npz \npf^{-1}$ from the left, one has
\begin{align}
  \dot{\yy} &= -\npz \npf^{-1} \npdd \npz^{-1} \yy +
  \npz \npf^{-1} \dri(\xx) + \npz \noi \nonumber \\
  &= -\npdd \npz^{-1} \yy +
  \npz \npf^{-1} \dri(\xx) + \npz \noi \ , \label{eq:notyet}
\end{align}
to first order in $\npd$. Before proceeding with the simplification
any further, one sees that the whole point of this exercise was to rid
the noise process of its perturbed prefactor, and make the
perturbation appear elsewhere.

Note the argument $\xx$, rather than $\yy$, of $\dri(\xx)$ in
\Eqref{eq:notyet}. In order to arrive at an equation of motion for the
coordinate $\yy$, we therefore have to expand:
\begin{align}
  \dri_i (\xx) &= \dri_i(\yy) + (\xx_j-\yy_j) \frac{\del
    \dri_i(\yy)}{\del \yy_j} \nonumber \\
  &= \dri_i (\yy) + \left(\npd \npz^{-1} \yy\right)_j 
  \frac{\del \dri_i(\yy)}{\del \yy_j} \ ,
\end{align}
where a summation is implied over repeated indices, and the difference
$\xx-\yy$ was given by \Eqref{eq:cov}. Noting also that $\npf^{-1} =
\npz^{-1} - \npz^{-1} \npd \npz^{-1}$ to first order, we can proceed
with the simplification of \Eqref{eq:notyet} to find
\begin{align}
  \dot{\yy}_i &=
  \left(1 - \npd \npz^{-1}\right)_{ij} \left[
    \dri_j (\yy) + \left(\npd \npz^{-1} \yy\right)_k 
  \frac{\del \dri_j(\yy)}{\del \yy_k} \right]
  -(\npdd \npz^{-1} \yy)_i 
  + \npz_{ij} \noi_j \ ,
\end{align}
or after discarding terms of order higher than 1 in $\npd$,
\begin{align}
  \dot{\yy}_i 
  &= \dri_i(\yy) - (\npd \npz^{-1})_{ij} \dri_j(\yy) +
  \left(\npd \npz^{-1} \yy\right)_k 
  \frac{\del \dri_i(\yy)}{\del \yy_k} 
  -(\npdd \npz^{-1} \yy)_i
  + \npz_{ij} \noi_j \ . \label{eq:almost}
\end{align}
We see that it is convenient to define
\begin{align}
  \hh (t) = \npd (t) \npz^{-1} \ , \label{eq:h}
\end{align}
which is a dimensionless smallness parameter matrix. After the
convenience of this parameter is duly exploited, it will be rewritten
in terms of the diffusivity rather than the noise prefactor.

Finally, we can rearrange \Eqref{eq:almost} into
\begin{align}
  \dot{\yy} = \driy(\yy) + \npz \noi \label{eq:eomy}
\end{align}
where
\begin{align}
  \driy (\yy) &= \dri (\yy) + \fp (\yy) \ , \\
  \fp_i (\yy) &= - \hh_{ik} \dri_k (\yy) +
  \hh_{kl} \yy_l \frac{\del \dri_i (\yy)} {\del \yy_k} - 
  \dot{\hh}_{ik} \yy_k \ , \label{eq:f1}
\end{align}
with $\dot{\hh} = \npdd \npz^{-1}$ according to
\Eqref{eq:h}. Thus, the perturbation of the noise prefactor in the
evolution of $\xx$ has been re-expressed as a perturbation of the mechanical 
force in the evolution of $\yy$. Ref.\ \cite{fal16b} discusses another
perspective from which this deformation of a thermal perturbation into
a deterministic perturbation can be viewed. Note that the
``pseudo-force'' $\fp (\yy(t);t)$ depends explicitly on time through
the perturbation $\hh (t)$.

\subsection{Path weight and response} \label{sec:resp}

The objective of linear response is to investigate the variation of
the expectation value of an observable to linear order in a small
perturbation. Taking the observable to be a state observable, 
$\obs (t) = \obs \vec( \xx (t) \vec)$, this
variation has the form
\begin{align}
  \delta \!\left< \obs(t) \right> = \left< \obs (t)
  \delta \ln \Pr [\xx] \right> \ , \label{eq:respon}
\end{align}
due simply to the variation of the probability weights, $\Pr [\xx]$,
of trajectories $\xx(t)$. The negative logarithm of the path weight,
\begin{align}
  - \ln \Pr [\xx] = \Act[\xx] \ ,
\end{align}
is commonly termed the path action. However, let us remark and remind
that due to technical difficulties of varying the action with respect
to noise amplitude, we will utilize the coordinate $\yy$ which, as
discussed earlier, reformulates the problem in terms of a variation of
the deterministic force.

According to the equation of motion \eqref{eq:eomy} for $\yy (t)$ with
the Gaussian noise process $\noi (t)$, the probability weight of a
trajectory $\yy(t)$ has the quadratic action
\begin{align}
  \Acty [\yy] = \int \frac {\dd s} 4 
  \left[ \dot{\yy}_i - \driy_i (\yy) \right]
  \difz^{-1}_{ij}
  \left[ \dot{\yy}_j - \driy_j (\yy) \right]
  + \int \frac {\dd s} 2 
  \frac{\del}{\del \yy_i} \driy_i (\yy) \label{eq:pw}
\end{align}
in the Stratonovich convention~\cite{zinn02} up to a trajectory-independent additive
constant, and the symmetric matrix $2 \difz = \npz
\npz^\dagger$.\footnote{For brevity of notation, we denote the
  elements of the inverse of a matrix as in $\dif^{-1}_{ij}$ rather
  than the unambiguous but cumbersome $(\dif^{-1})_{ij}$.} Also note
that evaluation of the integrands at time $s$ is implied.

The variation of \Eqref{eq:pw} to first order in $\fp (\yy)$ is
simply
\begin{align}
  \delta \Acty [\yy] = - \int \frac {\dd s} 2
  \left[ \difz^{-1}_{ij}
  \left( \dot{\yy}_j - \dri_j \right) 
  - \frac{\del } {\del \yy_i} \right] 
  \fp_i (\yy(s);s) \ . \label{eq:dlogP.1}
\end{align}
Note that due to its definition \eqref{eq:f1}, $\fp (\yy)$ is of at
least first order in the perturbation parameter $\hh (s)$. Thus, the
replacement $(\yy, \difz) \to (\xx, \dif)$ incurs errors only of
orders which are already truncated in this linear response
scheme. That is to say, in \Eqref{eq:respon} one can use $\delta \ln
\Pr [\xx] = -\delta \Act [\xx] = -\delta \Acty [\xx]$ via $(\yy,\difz)
\to (\xx,\dif)$ in \Eqref{eq:dlogP.1}. Using \Eqref{eq:f1} to express
$\fp (\xx)$ explicitly, we rewrite
\begin{alignat}{1}
  \delta \Act [\xx] = - \int \frac {\dd s}2 \left[ 
    \dif^{-1}_{ij} \left( \dot{\xx}_j - \dri_j \right) 
  - \del_i \right] 
  \left(- \hh_{ik} \dri_k + \hh_{kl} \xx_l \del_k \dri_i 
  - \dot{\hh}_{ik} \xx_k \right)  \ , 
\end{alignat}
where $\del_i$ is shorthand for $\del/\del x_i$. Here, one can count
powers of time derivative (dots on $\xx$ or $\hh$) to determine the
parity of each term under time-reversal, which is relevant for
thermodynamical arguments. Labeling according to the parity ($-/+$)
under time-reversal, we separate $\delta \Act [\xx]$
into\footnote{Note that $\del_i (\hh_{kl} \xx_l \del_k \dri_i -
  \hh_{ik} \dri_k) = \hh_{kl} \xx_l \del_k \del_i \dri_i + \hh_{kl}
  \delta_{il} \del_k \dri_i - \hh_{ik} \del_i \dri_k = \hh_{kl} \xx_l
  \del_k \del_i \dri_i$.}
\addtocounter{equation}{-1}
\begin{subequations}
\begin{align}
  \delta \Act_- [\xx] &= - \int \frac {\dd s}2 \left[
    \dif^{-1}_{ij} \dot{\xx}_j (\hh_{kl} \xx_l \del_k \dri_i
    -\hh_{ik} \dri_k ) 
    + \dot{\hh}_{ik} \xx_k \dif^{-1}_{ij} \dri_j 
    + \dot{\hh}_{kk} \right] \ , \label{eq:ent.1}
\end{align}
and
\begin{align}
  \delta \Act_+ [\xx] = - \int \frac {\dd s}2  
    \dif^{-1}_{ij} \dri_j  
  (\hh_{ik} \dri_k - \hh_{kl} \xx_l \del_k \dri_i) 
   + \int \frac {\dd s}2 
  (\hh_{kl} \xx_l \del_k \del_i \dri_i 
   +\dif^{-1}_{ij} \dot{\xx}_j \dot{\hh}_{ik} \xx_k) \ ,
   \label{eq:fre.1}
\end{align}
\end{subequations}
with $\delta \Act [\xx] = \delta \Act_+ [\xx] + \delta \Act_- [\xx]$.
Via integration by parts, $\dot{\hh}(s)$ can be exchanged with
$\hh(s)$.\footnote{For this matter, we note here that we have
  intentionally left out integration limits in the action: The forcing
  $\hh (s)$ is assumed to be temporally localized, and the integration
  domain is infinite in principle. Thus, $\int \dd s\, \dot{\hh} g =
  -\int \dd s\, \hh \dot{g}$ for a generic function $g(s)$, and
  likewise $\int \dd s\, \dot{\hh}_{kk} = 0$ in \Eqref{eq:ent.1}.}
After also manipulating the indices for the convenience of isolating a
common factor, $\hh_{kl}(s)$, one has
\begin{subequations}
\begin{align}
  \delta \Act_- [\xx] &= - \int \frac {\dd s}2 \left[
    \dif^{-1}_{ij} \dot{\xx}_j \xx_l \del_k \dri_i 
    -\dif^{-1}_{kj} (\dot{\xx}_j \dri_l + \dot{\xx}_l \dri_j)
    - \dif^{-1}_{kj} \dot{\xx}_i \xx_l \del_i \dri_j
    \right] \hh_{kl} \ , \label{eq:ent.2}\\
  \delta \Act_+ [\xx] &= - \int \frac {\dd s}2 \left[
    \dif^{-1}_{kj} \dri_j \dri_l 
    - \dif^{-1}_{ij} \dri_j \xx_l \del_k \dri_i 
    - \xx_l \del_k \del_i \dri_i
    + \dif^{-1}_{kj} \tfrac{\dd}{\dd s} (\dot{\xx}_j \xx_l)
    \right] \hh_{kl} \ , \label{eq:fre.2} 
\end{align}
\end{subequations}
where the identity $\dot{\dri}_j = \dot{\xx}_i \del_i \dri_j$ was used
in $(\dd/\dd s) (\xx_l \dri_j) = \dot{\xx}_l \dri_j + \xx_l
\dot{\dri}_j$, owing to Stratonovich calculus, concerning the very
last term.

Up until now, we have not elaborated about the dimensionless parameter
matrix $\hh (s)$, which we do in Appendix~\ref{app:h}. The gist is that, rather
than expressing $\hh (s)$ via the somewhat nonphysical noise prefactor
$\npf$ via \Eqref{eq:h}, one can write it in terms of the variation of
the inverse diffusivity, $\delta \dif^{-1}$, which is a more sensible
physical quantity. The manifest symmetry in the contents of the
resulting expression, $\hh = (-1/2)\dif \delta \dif^{-1}$,
facilitates certain simplifications on $\delta \Act [\xx]$. In
particular, now that $\dif^{-1}_{jk} \hh_{kl} = (-1/2) \delta
\dif^{-1}_{jl}$ is a symmetric matrix, the last term of
\Eqref{eq:fre.2} can benefit from
\begin{align}
  \dif^{-1}_{kj} \hh_{kl} \tfrac{\dd}{\dd s} (\dot{\xx}_j \xx_l) 
  =  -\tfrac 12 \delta \dif^{-1}_{jl} 
  \tfrac{\dd}{\dd s} (\dot{\xx}_j \xx_l)
  = -\tfrac 14 \delta \dif^{-1}_{jl}
  \tfrac{\dd^2}{\dd s^2} (\xx_j \xx_l) \ .
\end{align}
Similarly, the two middle terms of \Eqref{eq:ent.2} combine. With
these simplifications we finally have
\begin{subequations}
\begin{align}
  \delta \Act_- [\xx] &= \int \frac {\dd s \, \dot{\xx}_i}4 
  \left[ \left( \dif^{-1}_{ij} \dif_{km} - \delta_{ik} \delta_{jm} \right)
    \xx_l \del_k \dri_j - 2 \delta_{im} \dri_l 
    \right]  \delta \dif^{-1}_{ml} \ , \label{eq:cor.ent}\\
  \delta \Act_+ [\xx] &= \int \frac {\dd s}4 
  \left[ \dri_m \dri_l - \dif_{km} \xx_l
    \left( \dif^{-1}_{ij} \dri_j \del_k \dri_i + \del_k \del_i \dri_i
    \right) + \tfrac 12 \tfrac{\dd^2}{\dd s^2} (\xx_m \xx_l ) 
    \right] \delta \dif^{-1}_{ml} \ . \label{eq:cor.fre} 
\end{align}
\end{subequations}
At this stage, it would be possible to define a tensorial response
function for a state observable $\obs (t) = \obs \vec(\xx(t)\vec)$ to
an impulsive variation in the inverse diffusivity matrix as
\begin{align}
  \RF_{lm} (t,s) &= -\left< \obs \vec(\xx(t)\vec)
  \frac {\delta \Act [\xx]} {\delta \dif^{-1}_{lm} (s)} 
  \right> \ ,
\end{align}
which satisfies $\delta \!\left< \obs (t) \right> = -\left< \obs(t)
\delta \Act [\xx] \right> = \int \dd s \, \RF_{lm} (t,s) \delta
\dif^{-1}_{lm}(s)$, yielding
\begin{align}
\RF_{lm} (t,s) \equiv \RF_{lm}^- (t,s) + \RF_{lm}^+(t,s)
\end{align}
with
\addtocounter{equation}{-1}
\begin{subequations}\label{eq:R}
\begin{align} 
  \RF_{lm}^- (t,s) &= -\left< \frac {\obs(t)}4 
    \left[ \left( \dif^{-1}_{ij} \dif_{km} 
    - \delta_{ik} \delta_{jm} \right) \dot{\xx}_i
    \xx_l \del_k \dri_j - 2 \dot{\xx}_m \dri_l 
    \right]\!(s) \right> \ , \label{eq:RF.ent} \\
  \RF_{lm}^+ (t,s) &= -\left< \frac {\obs(t)}4 
  \left[ \dri_m \dri_l - \dif_{km} \xx_l
    \left( \dif^{-1}_{ij} \dri_j \del_k \dri_i 
    + \del_k \del_i \dri_i \right) 
    + \tfrac 12 \tfrac{\dd^2}{\dd s^2} (\xx_m \xx_l ) 
    \right]\!(s) \right> \ . \label{eq:RF.fre}
\end{align}
\end{subequations}
 These expressions are the generalization of those of
Refs.\ \cite{fal16,fal16b} to non-negligible state-independent
hydrodynamic coupling. Note the ``collective'' argument $(s)$ that
applies to all the time dependence inside the square brackets
preceding it.

\subsection{Stationary susceptibility} \label{sec:sus}

The integrated response is called susceptibility. 
In practice, one is usually able to manipulate a single scalar
parameter, call it $\prm (s)$. It suits
to write the variation $\delta \dif^{-1}_{lm} (s)$ in terms of the
variation $\delta \prm (s)$ of the parameter (to first order): 
$\delta\dif^{-1}_{lm} (s) = \delta \prm (s) (\del \dif^{-1}_{lm} /\del \prm)$. 
Then, an empirically accessible susceptibility to a stepwise
change in $\prm$ at $t=0$ can be expressed as
\begin{align}
  \sus (t) &=
  \left.\frac{\del\mean{\obs(t)}^\prm}{\del \prm}\right|_{\prm=\prm(0)}
  = \int_0^t \dd s \, \HH_{lm} \RF_{lm} (t,s) \ , \label{eq:sus} 
\end{align}
with 
\begin{align}
  \label{eq:H}
  \HH_{lm} &\equiv \left. \frac {\del \dif^{-1}_{lm}} {\del \prm}\right|_{\prm = \prm(0)} \, .
\end{align}

The time integral of \Eqref{eq:sus} removes one of the derivatives
$\dd/\dd s$ in the last term of \Eqref{eq:RF.fre} and gives the
boundary term,
\begin{align} {\textstyle
  -\left< \obs(t) \tfrac 18 \tfrac {\dd}{\dd s} 
  [ \xx_m (s) \xx_l (s)] \big|_{0}^{t} \right>
   \ .}
\end{align}
The remaining derivative can also be removed if we specialize to the case
where the initial distribution $\dist \vec( \xx(0) \vec)$ in the
unperturbed averages $\left< \cdot \right>$ is stationary, since, due
to invariance under time translation, the response function becomes
$\RF (t,s) = \RF (t-s)$, and in particular
\begin{align}
  -\left< \obs (t) \tfrac 18 \tfrac{\dd}{\dd s} 
  [ \xx_m (s) \xx_l(s) ] \big|_{0}^{t} \right>
  = \left< \tfrac 18 \tfrac{\dd \obs(t)}{\dd t} 
  [ \xx_m (s) \xx_l (s) ] \big|_{0}^{t}\right>  \ .
\end{align}
The time derivative at the later instant $t$ can then be replaced by
the backward generator of the dynamics, \ie.,
\begin{align}
  \left< \tfrac 18 \tfrac{\dd \obs(t)}{\dd t} 
  [ \xx_m (s) \xx_l (s) ] \big|_{0}^{t}\right> =
  \left< \tfrac 18 \Lgen \obs(t) 
  [ \xx_m (s) \xx_l (s) ] \big|_{0}^{t}\right> \ , \label{eq:dtL}
\end{align}
where
\begin{align}
  \Lgen = \dri_i \del_i + \dif_{ij} \del_i \del_j \ . \label{eq:Lgen}
\end{align}
This replacement is desirable, because the original expression on the
left-hand side of \Eqref{eq:dtL} with the time derivative is much
noisier, and thus requires much more statistics to converge.  Since
our application will be in a the stationary state, we take advantage
of this and continue from here on with the assumption of stationarity.

As previously done~\cite{fal16b}, we split the susceptibility into four terms
\begin{align}\label{eq:suscep}
  \sus (t) &= 
  \sus^{-}_1 (t) + \sus^{-}_2 (t) +
  \sus^{+}_1 (t) + \sus^{+}_2 (t) \ ,
\end{align}
where
\addtocounter{equation}{-1}
\begin{subequations}
\label{eq:sus4terms}
\begin{align}
  \sus^{-}_1 (t) &= \frac 1 2 \int_0^t \dd s  \,
  \HH_{ml} \mean{\obs(t)[\dot{\xx}_m \dri_l ](s) }
  \, , \label{eq:sus.ent1}\\
  \sus^{-}_2 (t) &= \frac 1 4  \int_0^t \dd s  \,
  \HH_{ml} \mean{\obs(t) 
    [ (\delta_{ik} \delta_{jm} -\dif^{-1}_{ij} \dif_{km} )
      \dot{\xx}_i \xx_l \del_k \dri_j ](s) } 
  \, ,    \label{eq:sus.ent2}\\
  \sus^{+}_1 (t) &= \frac 1 4  \int_0^t \dd s \,
  \HH_{ml} \mean{ \obs(t) 
    [ \xx_l \dif_{km} \dif^{-1}_{ij} \dri_j \del_k \dri_i
      -\dri_m \dri_l ](s) }
  \, , \label{eq:sus.fre1}\\
  \sus^{+}_2 (t) &=   \frac 1 8  \HH_{ml} 
  \mean{ \Lgen \obs(t)\, \xx_m (s) \xx_l (s) \big|_{0}^{t} }
  \, . \label{eq:sus.fre2}
\end{align}
\end{subequations}
are written in terms of Stratonovich stochastic integrals.
If the heat bath is in equilibrium, the term $\sus^{-}_1 (t)$
corresponds to the correlation between the observable and half of the
entropy production in the bath. This interpretation, previously discussed for
the linear response to mechanical forces~\cite{bai09,bai09b} and
temperature variations~\cite{bai14,fal16,fal16b},
is not possible in our experiment and we remain with the fact that $\sus^{-}_1 (t)$ as
well as $\sus^{-}_2(t)$ correlate the observable with
time-antisymmetric, \ie., entropy-production-like quantities.  The
correlations in $\sus^{+}_1 (t)$ and $\sus^{+}_2 (t)$ instead involve
time-symmetric, non-dissipative variables. They were named
``frenetic'' terms because often those time-symmetric quantities are a
measure of how frenzy the system's behaviour in state space is
\cite{bai09,bai09b,bai13}.

\subsection{Susceptibility in matrix notation} \label{sec:sus_mat}
 
Despite the many indices appearing in \Eqref{eq:sus4terms}, note that
the highest rank tensor involved above is still of rank two.
Therefore, the index contractions can be broken up into ordinary
matrix operations.  This should be useful, for example, for a
numerical implementation in a software where matrix multiplication is
defined.

We denote by $\nabla \dri^\dagger$ the matrix with elements 
$\del_i \dri_j$ (where $i$ is the row index).
Thus, we deal with vectors (the positions $\xx$, velocities $\dot \xx$,
and drifts $\dri$) and matrices. The latter are
the noise variation
$\HH$ (defined in \Eqref{eq:H}), the diffusivity matrix $\dif$ and
its inverse $\dif^{-1}$, and  
$\nabla \dri^\dagger$.  By
simply following the sequence of index repetitions, it is
straightforward to express the terms in the susceptibility as
\begin{subequations}
\label{eq:sus.matr}
\begin{align}
  \sus^{-}_1 (t) &= \frac 1 2 { \int_0^t} \dd s \,
  \mean{ \obs(t) [\dot{\xx}^\dagger \HH \dri](s)}
    \ , \label{eq:sus.ent.m1}\\
  \sus^{-}_2 (t) &= \frac 1 4 { \int_0^t} \dd s  \,
  \left< \obs(t) [
     \dot{\xx}^\dagger (\nabla \dri^\dagger) \HH \xx 
    - \xx^\dagger \HH \dif (\nabla \dri^\dagger) \dif^{-1} \dot{\xx}
    ](s) \right> 
    \ , \label{eq:sus.ent.m2}\\
  \sus^{+}_1 (t) &= 
     \frac 1 4  {\int_0^t}\dd s \,
  \left< \obs(t) [\xx^\dagger \HH \dif (\nabla \dri^\dagger)  
          \dif^{-1} \dri - \dri^\dagger \HH \dri ](s) \right> 
  \ , \label{eq:sus.fre.m1}\\
  \sus^{+}_2 (t) &= 
  \frac 18 \mean{\Lgen \obs(t)\, [\xx^\dagger \HH \xx ](s)} \big|_{s=0}^{s=t}
  \ . \label{eq:sus.fre.m2}
\end{align}
If the stationary conditions are not satisfied, the last term should
be evaluated as 
\begin{align}
\sus^{+}_2(t)= - \frac 18 \frac {\dd}{\dd s}\mean{\obs(t)\,[\xx^\dagger\HH \xx](s)}\big|_{s=0}^{s=t} \,.
\end{align}
\end{subequations}

\section{Susceptibility of the colloidal system} \label{sec:sus_coll}

We calculate the susceptibility of the
colloidal system to a change in the effective temperature 
of the stochastic driving. We thus identify $\prm =\Tem$ and
with \Eqref{eq:dif.2} we find
\begin{align}
  \HH = \frac {\del (\dif^{-1})} {\del \Tem} = \begin{bmatrix}
    \frac {-\drag} {(T + \Tem)^2} & 0 \\0 &0 \end{bmatrix} \ .
\end{align}
Using this expression in \Eqref{eq:sus} via the response functions
\eqref{eq:RF.ent}--\eqref{eq:RF.fre} along with the simplifications
sketched above, we find the susceptibility.  Note that the term of the
time-symmetric part \eqref{eq:RF.fre} proportional to $\del_k \del_i
\dri_i$, while in general nonzero, disappears for the particular
system in question, since two derivatives of the mean drift 
$\dri =\mob \for = -\mob \kappa \xx$ vanish due to the quadratic potential
\eqref{eq:U}.

Experimental data were sampled at constant rate, with a time step
$\Delta t$.  This means that a trajectory is a discrete collection of
points $x(0),x(1), x(2),\ldots,x(n)$ with $n = t /\Delta t$.  For each
temperature we have a long time series composed of $n_s = 10^6$
samples, out of which we extract $n_s-n+1$ trajectories by shifting
the initial point. Thus, these trajectories are partially overlapped
with other ones.  Since the autocorrelation time of the system is
about  $\gamma/\kappa_{11}\approx \gamma/\kappa_{22}\approx 5$ms 
$=4 \Delta t$, we have taken the correlation of
data into account by amplifying the error bars in our
results (standard deviations) by a factor of $\sqrt{4} = 2$.

Velocities are defined as $\dot x(i) = [x(i+1)-x(i)]/\Delta t$.  The
integrals in \Eqref{eq:sus4terms} or \Eqref{eq:sus.matr} follow the
Stratonovich convention.  In their time-discretized version,
quantities (other than the $\dot x$ mentioned above) in the integrals
are averaged over time steps. For example, the average
$[x(i+1)+x(i)]/2$ represents the positions at time step $i$.

\begin{figure}[!tb]
\begin{tabular}{cc}
$\Tem = 0$ K
&
$\Tem = 120$ K \\
\includegraphics[width=7.4cm]{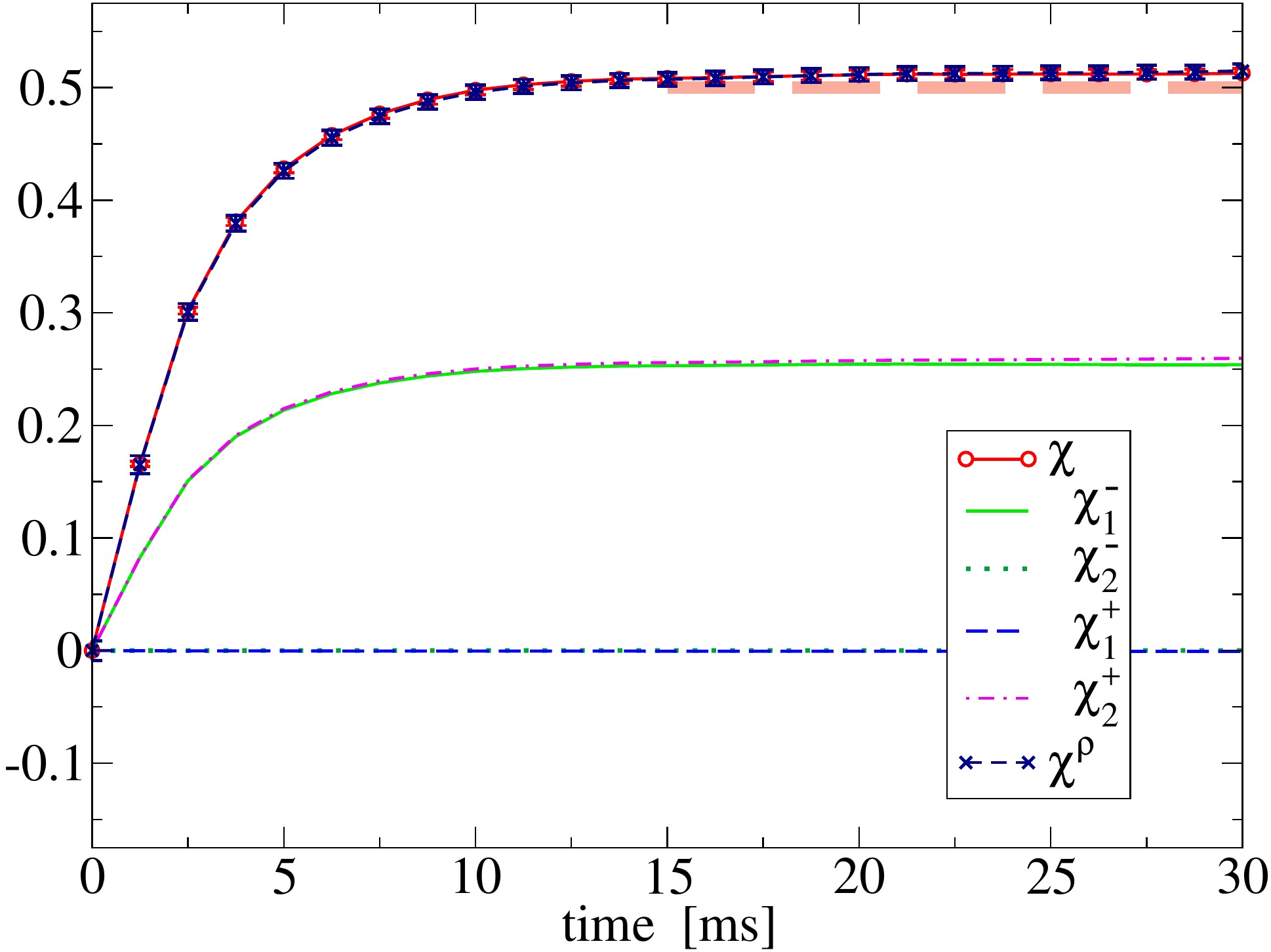}
&
\includegraphics[width=7.4cm]{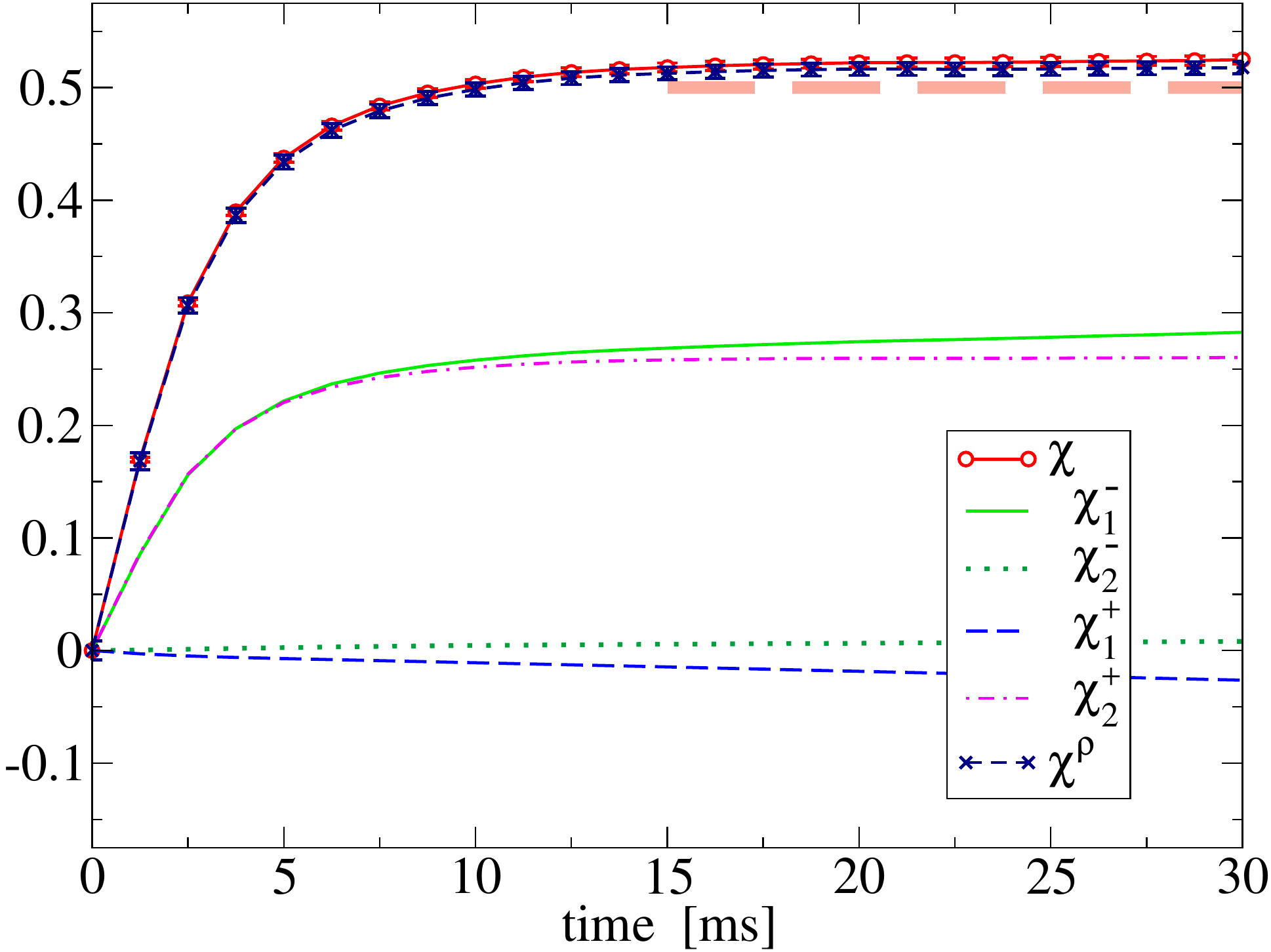}\\
$\Tem = 340.8$ K
&
$\Tem = 967.8$ K\\
\includegraphics[width=7.4cm]{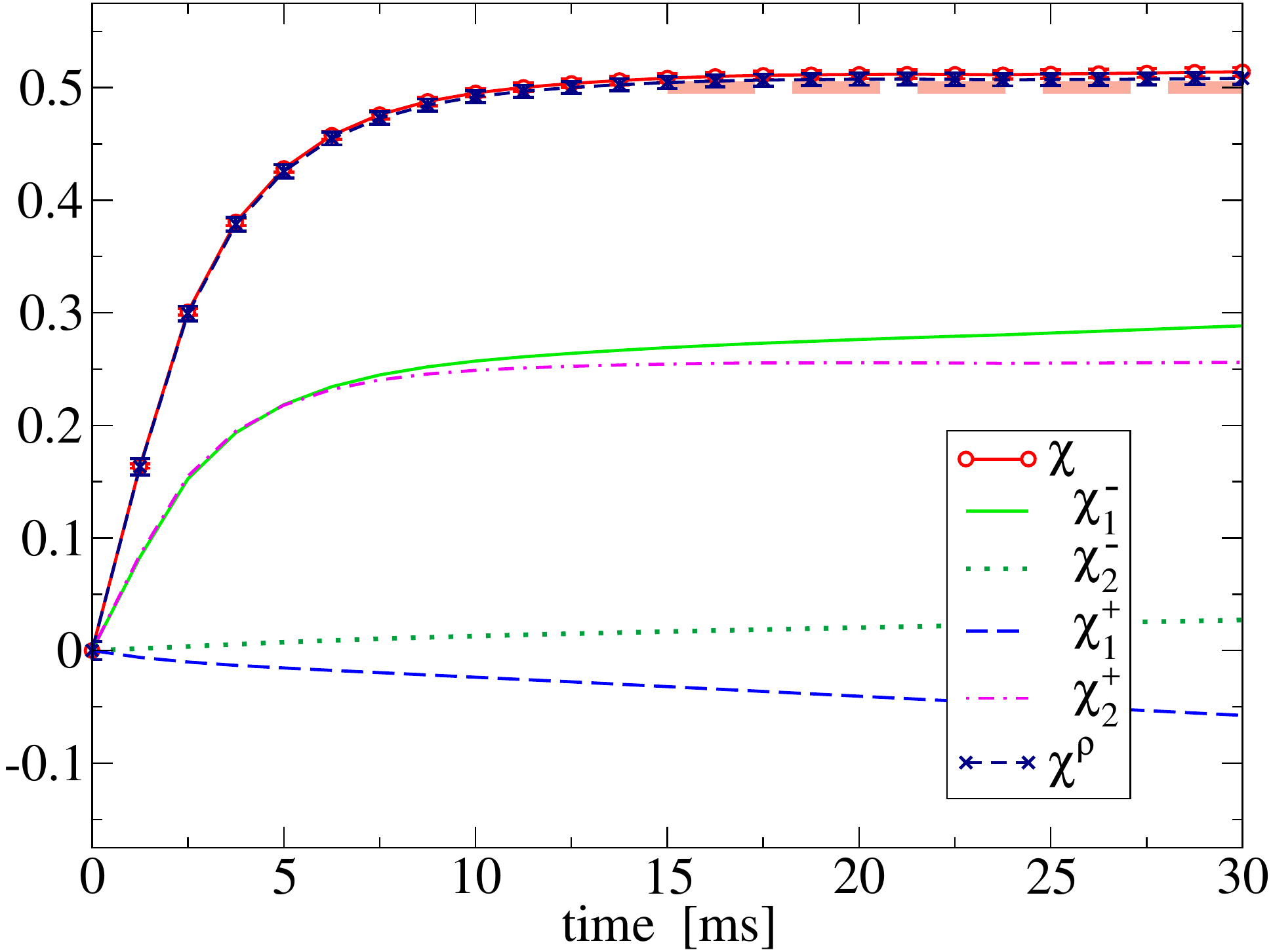}
&
\includegraphics[width=7.4cm]{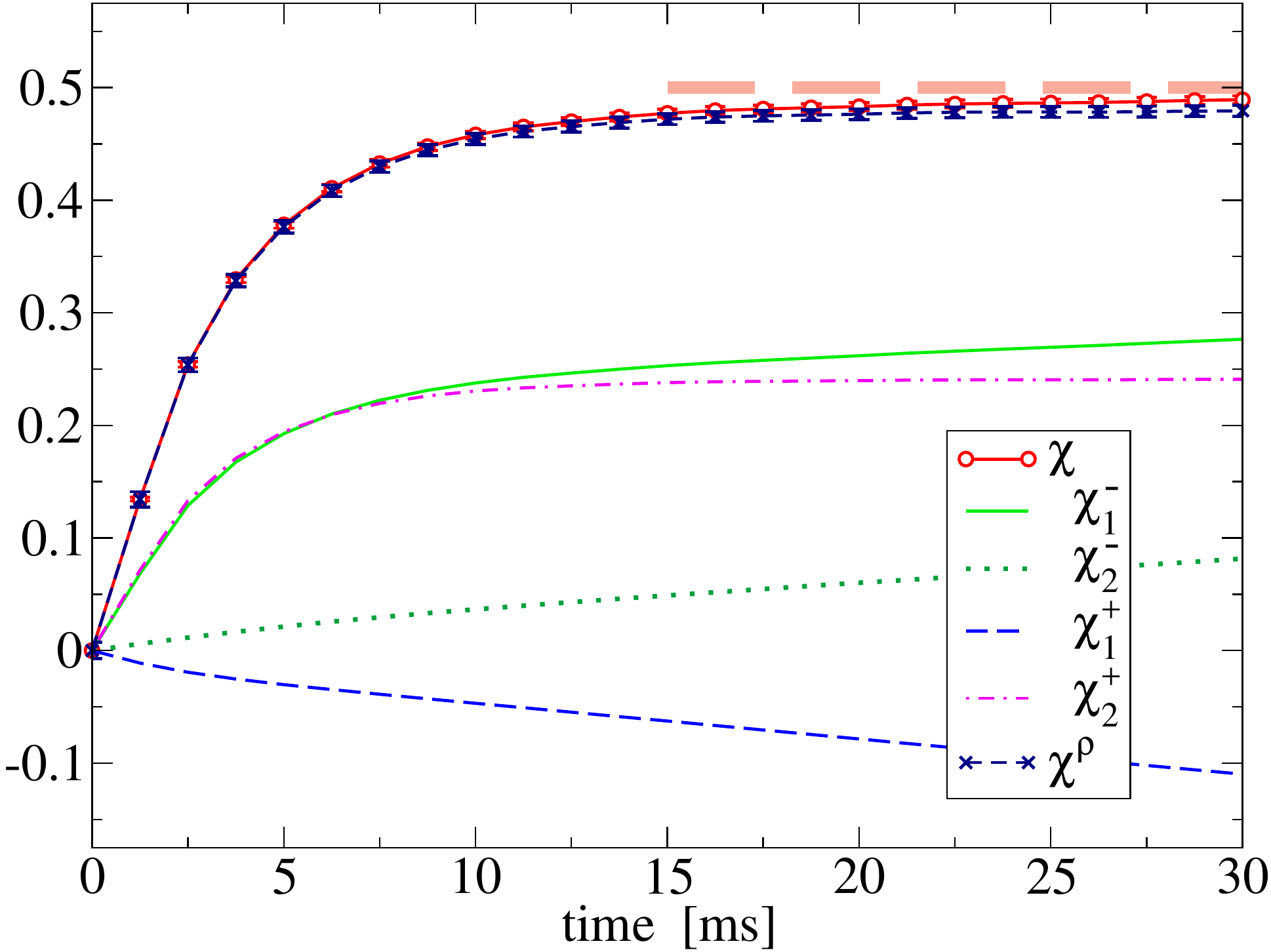}
\end{tabular}
\caption{Response of the total energy $U$ to a stepwise change of
  $\Tem$, for equilibrium ($\Tem = 0\,$K) and for
  nonequilibrium ($\Tem=120\,$K, $\Tem=340.8\,$K, and
  $\Tem=967.8\,$K). The susceptibility $\sus$ computed with the
  fluctuation-response relation is shown, together with its components
  $\sus^-_1$, $\sus^-_2$, $\sus^+_1$, and $\sus^+_2$.  The
  susceptibility $\chi^\rho$ computed using the response formula of
  Seifert and Speck \cite{sei10} is also shown.  
  The thick horizontal (pink) dashed line indicates the 
  asymptotic value, which is determined analytically in Appendix~\ref{app:asy}.
}
\label{fig:U}
\end{figure}

To visualize the linear response of some observables, we start by
considering the total energy $U = \frac{\kappa_{11}}2 x_1^2 +
\frac{\kappa_{22}}2 x_2^2$.  The susceptibility of the energy can be
considered a generalized heat capacity.  In Figure~\ref{fig:U} this
susceptibility $\sus$ is plotted as a function of time for four cases:
$\Tem=0$~K, which is the equilibrium condition with no forcing on particle $1$, 
and $\Tem=120$~K, $340.8$~K, and $\Tem=967.8$~K, 
which become progressively farther from equilibrium.  
In equilibrium one can notice that twice the
entropic term $\sus^-_1$ (which by definition is half of the Kubo
formula for equilibrium systems), would be sufficient to obtain the
susceptibility.  This picture breaks down out of equilibrium, where it
is the whole sum of the four components that yields the correct form
of the susceptibility.  For comparison, we show also the estimate
$\chi^\rho(t)$ obtained with the approach by Seifert and Speck
\cite{sei10}, through
\begin{align}
    \chi^\rho (t) = \left< \Obs(t) \left[
    \frac{\del \ln \rho (\xx(t)) }{\del \Tem} 
    - \frac{\del \ln \rho (\xx(0))}{\del \Tem} 
    \right] \right> \ ,
\end{align}
which is an integrated version of their response formula.
Appendix~\ref{app:st} sketches how the stationary distribution $\rho(\xx)$ for
this system was computed. Figure~\ref{fig:U} shows that
the susceptibilities evaluated with the two
approaches are compatible with each other.

\begin{figure}[!t]
\begin{tabular}{cc}
$\Tem = 0$ K
&
$\Tem = 120$ K \\
\includegraphics[width=7.4cm]{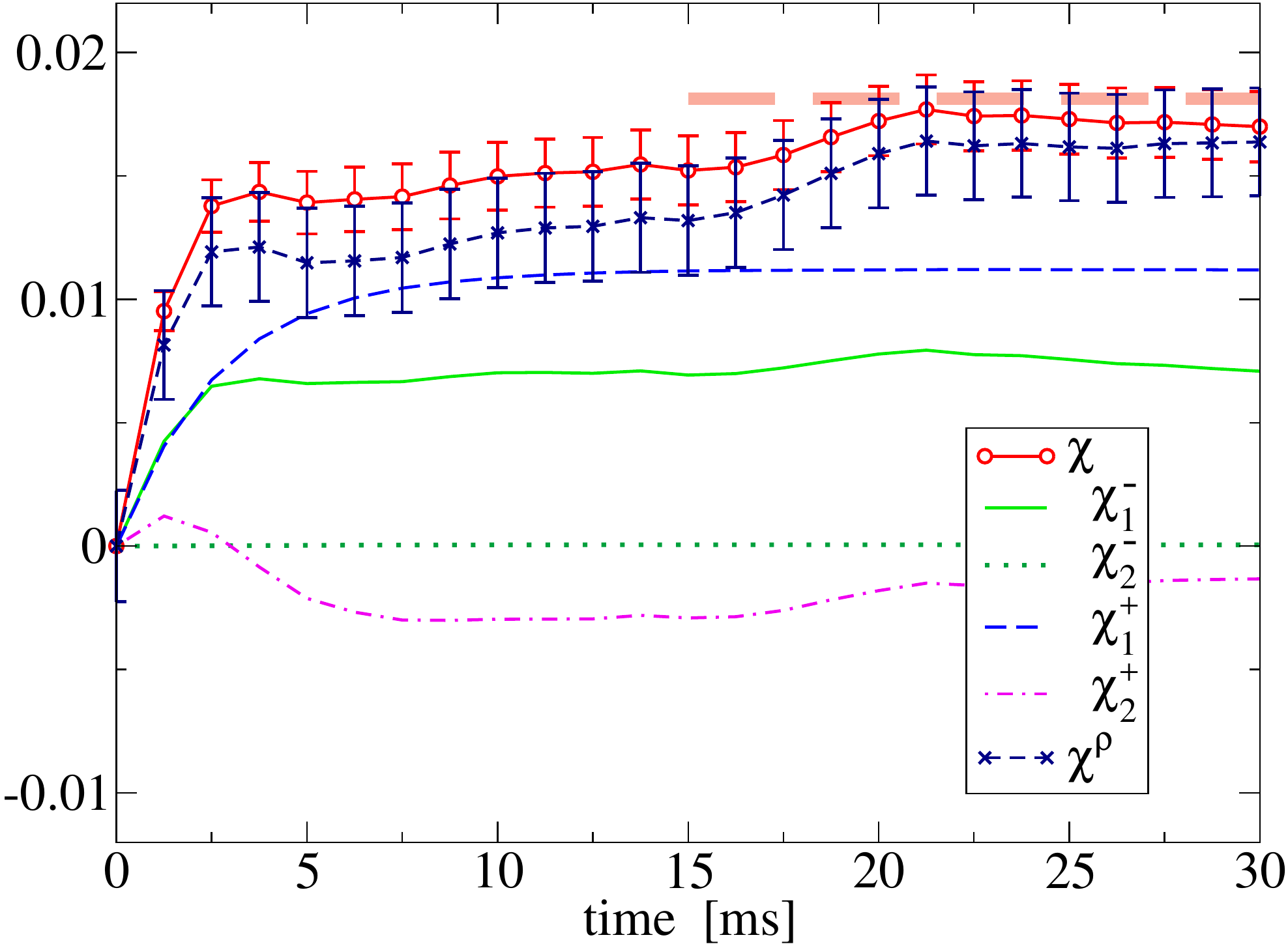}
&
\includegraphics[width=7.4cm]{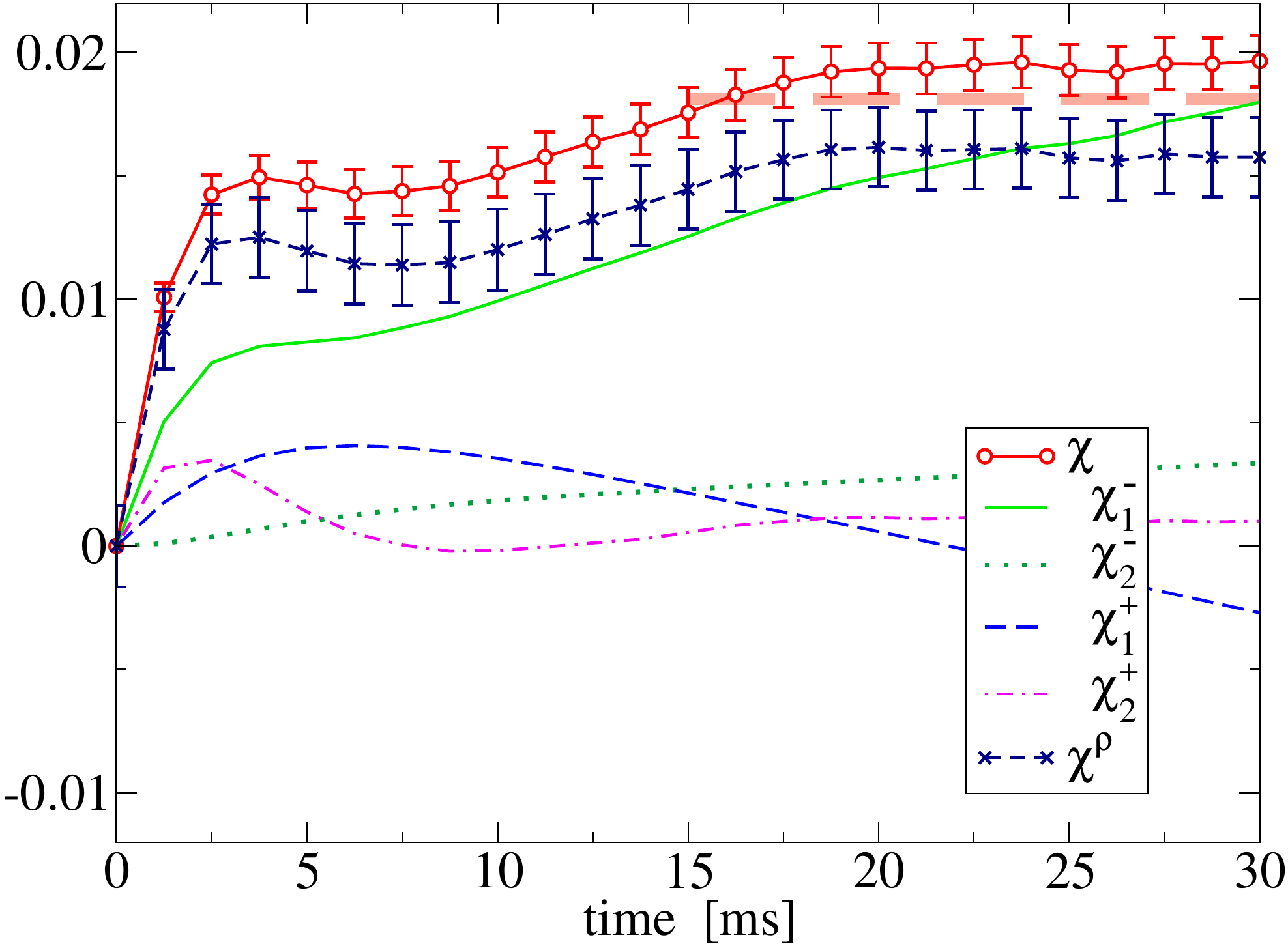}\\
$\Tem = 340.8$ K
&
$\Tem = 967.8$ K\\
\includegraphics[width=7.4cm]{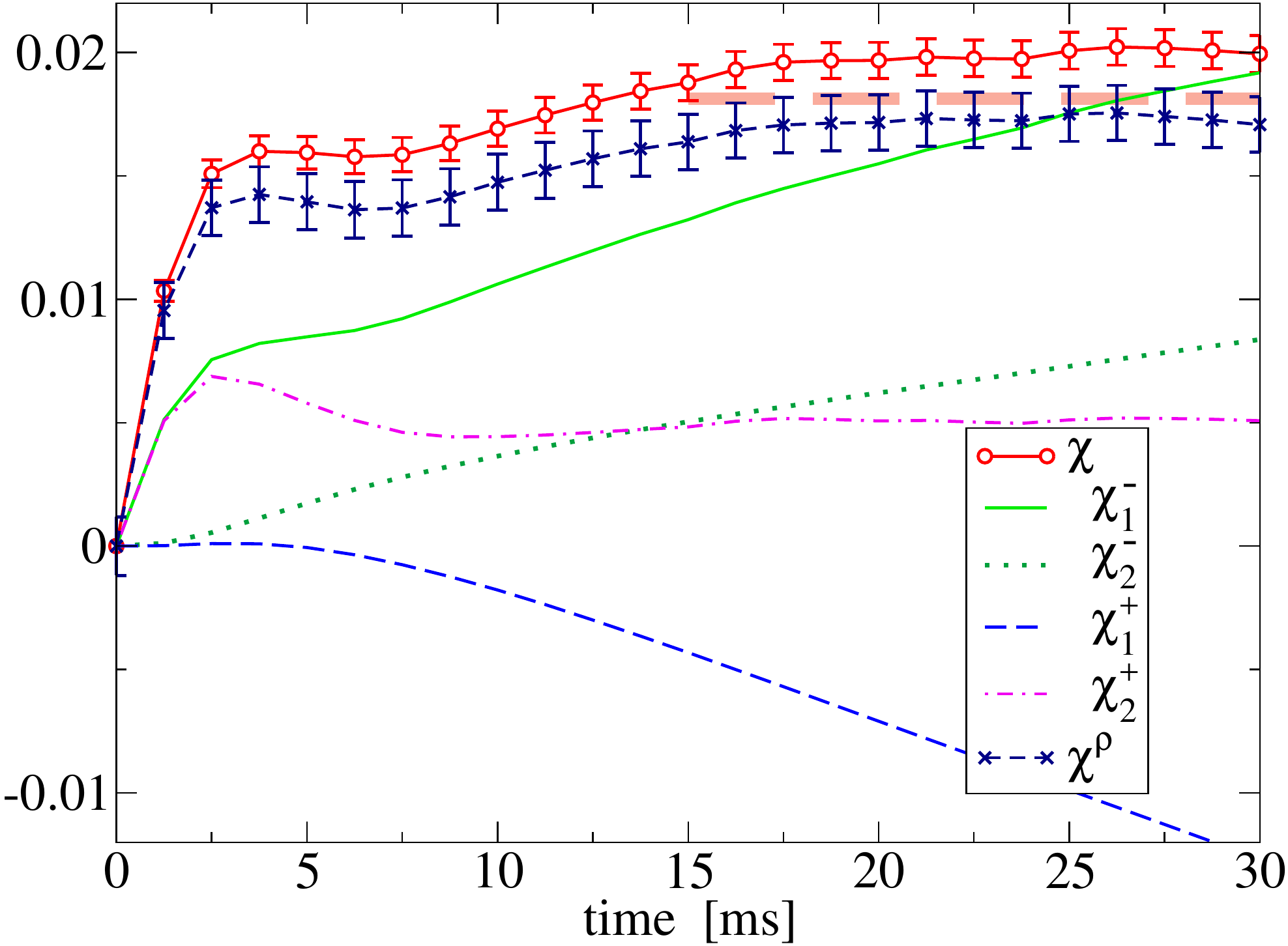}
&
\includegraphics[width=7.4cm]{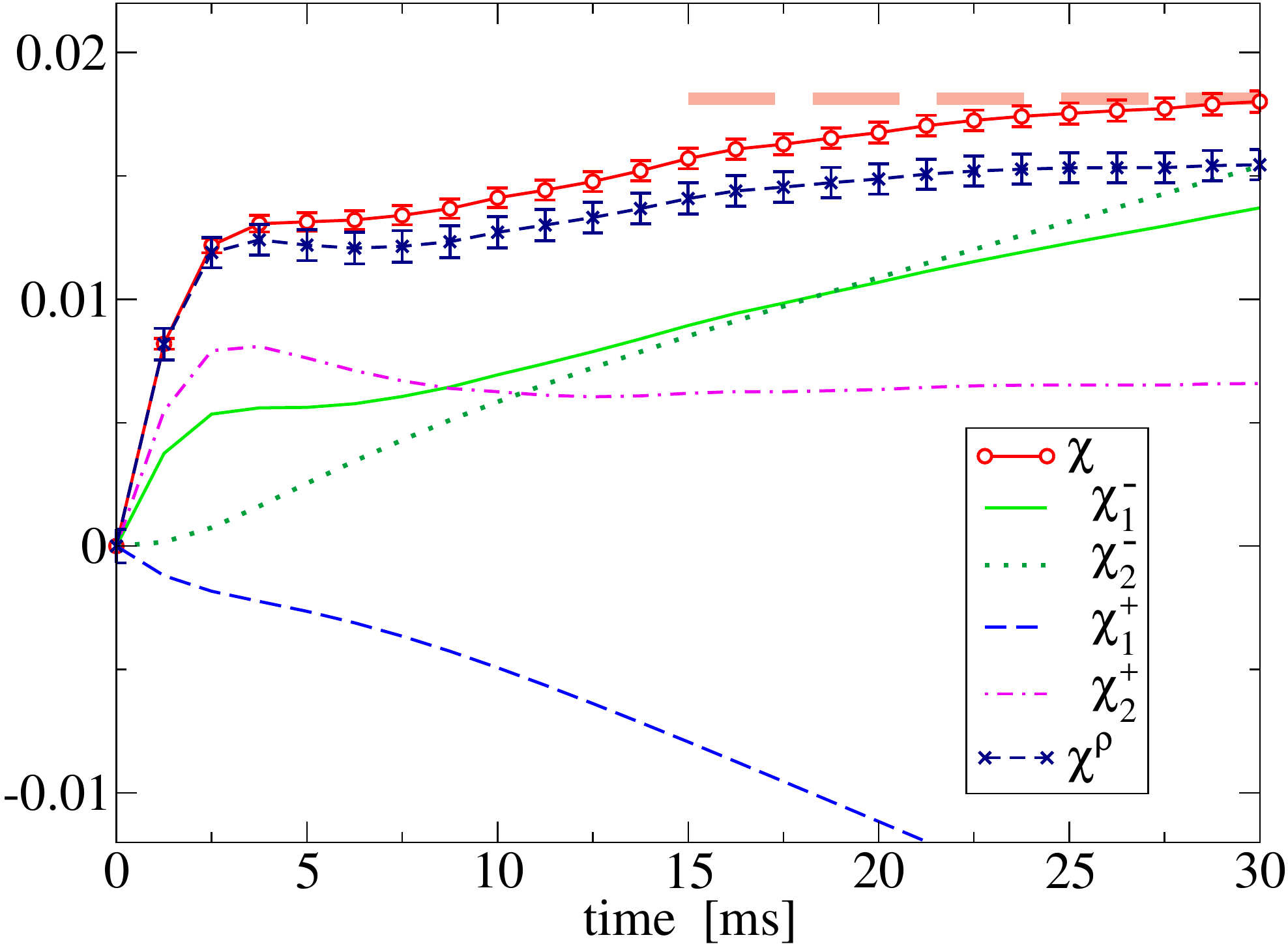}
\end{tabular}
\caption{Response of the energy $U_2$ of the second particle to a
  stepwise change of $\Tem$, see the legend of 
  Figure~\ref{fig:U} for all the details.}
\label{fig:U2}
\end{figure}

As a second example, we show the susceptibility to a change of $\Tem$
of the energy $U_2 = \frac{\kappa_{22}}2 x_2^2$ of the second particle,
namely the colloid which is not experiencing the random forcing directly.  
Yet, due to the hydrodynamic interaction, its energy on average increases with 
$\Tem$, as manifested by the susceptibilities plotted in Figure~\ref{fig:U2}.
Again, there is a fair agreement between the two estimates.  The
comparison with previous numerical estimates \cite{yol16}, which gave
very similar curves but for other values of parameters, corroborates
the validity of these results.


\section*{Conclusions}

We have proposed a quite general formalism for estimating the linear
response of an overdamped stochastic system to a variation of its
(additive) white-noise amplitude. This is a useful tool
for analyzing the response of experimental data, as for the discussed
system of two hydrodynamically coupled colloids in optical traps,
for which we are able to compute the susceptibility to a change
of an external random driving that reproduces a sort of higher
temperature for only one of the colloids.
Yet, note that our formalism if quite flexible and can be applied to
more abstract stochastic diffusive dynamics with correlated noise.
The next evolution of this approach should try to complete the picture
by dealing with systems with multiplicative noise.
Also the thermal response for nonequilibrium inertial systems should
be developed without resorting to the time-discretization used in 
a previous work~\cite{bai14}. This would allow, for example,
to determine nonequilibrium steady-state thermal susceptibilities of 
the Fermi-Pasta-Ulam model~\cite{lep03,dha08} or other
Hamiltonian models of heat conduction.

Similarly to the linear response of nonequilibrium systems to
mechanical perturbations, we note that the susceptibility to
temperature changes arises from dissipative as well as non-dissipative
aspects of the dynamics. This differs from the simple structure of
response theory for equilibrium systems, where only correlations
between heat fluxes (i.e., dissipation) and observable are
relevant for computing susceptibilities.
With these considerations in mind, the kind of 
fluctuation-response relations we discussed
might be used to better understand and quantify
the thermal response of nonequilibrium systems
such as relaxing glasses~\cite{bar99,lip07,cam09,mag10,cri13}
and active matter~\cite{mag10,loi11,gau12,mar13}.


\appendix

\section*{Appendices}

\section{Stationary distribution} \label{app:st}
Here we briefly sketch the calculation of the stationary distribution
of the system, for the sake of completeness.  

With a drift $\dri(\xx)=-\mob \kappa \xx$ linear in $\xx$, the
equation of motion \eqref{eq:eom} describes an Ornstein-Uhlenbeck
process, which we rewrite here as
\begin{align}
  \dot{\xx} = - \Omg \xx + \sqrt{2\dif} \noi \ ,
\end{align}
with $\Omg = \mob \kappa$. According to the corresponding
Fokker-Planck equation, $\dot{\rho} = -\del_i J_i$, stationarity
requires the ensemble current $J = -\rho \Omg \xx - \dif \nabla
\rho$ to be divergenceless, yielding
\begin{align}
  0 = \rho \Omg_{ii} + \Omg_{ij} \xx_j \del_i \rho
  +\dif_{ij} \del_i \del_j \rho \ .
\end{align}
Clearly, a Gaussian distribution of the form
\begin{align}
  \rho (\xx) = Z^{-1} \ee^{-\frac 12 \xx^\dagger \GG \xx} \label{eq:stdist}
\end{align}
satisfies the zero divergence condition above. To find the inverse
covariance matrix $\GG$, one substitutes the ansatz, upon which a few
lines of algebra implies the condition
\begin{align}
  \Omg \GG^{-1} + \GG^{-1} \Omg^\dagger = 2 \dif \ .
\end{align}
This is a system of linear equations in the elements of the covariance
matrix $\GG^{-1}$, which can be solved, for instance, by rewriting it
in terms of a vector composed of the columns of $\GG^{-1}$. Here, we
simply quote the resulting matrix: Recalling $\Omg = \mob \kappa$ with
$\kappa$ diagonal and $\mob$ given in \Eqref{eq:mob}, and $\dif$ given
in \Eqref{eq:dif.2}, one finds
\begin{align}
  \GG^{-1} = \begin{bmatrix} 
    \frac{T+(1-\varepsilon^2)\Tem}{\kappa_{11}} +
    \frac{\varepsilon^2\Tem}{\kappa_{11}+\kappa_{22}} &
    \frac {\varepsilon \Tem}{\kappa_{11} + \kappa_{22}} \\
    \frac {\varepsilon \Tem}{\kappa_{11} + \kappa_{22}} &
    \frac {T}{\kappa_{22}} + 
    \frac {\varepsilon^2\Tem}{\kappa_{11}+\kappa_{22}}
    \end{bmatrix} \ .
\end{align}
Inversion of this matrix thus determines the stationary distribution
\eqref{eq:stdist}.

\section{Form of the perturbation parameter} \label{app:h}
Even though it may sometimes be more convenient mathematically to work
in terms of the noise amplitude $\npf$, it is the diffusivity matrix
$\dif$ which is physically relevant, since the noise amplitude is
fixed only up to an orthogonal transformation. Hence the perturbation
parameter, $\hh = \npd \npz^{-1}$, defined in Sec.\ \ref{sec:cov} in
terms of the noise amplitude should eventually be expressed in terms
of a perturbation of the diffusivity. With $\dif = \difz + \delta
\dif$ and $\npf = \npz + \npd$, in line with Sec.\ \ref{sec:cov}, the
definitions $2 \dif = \npf \npf^\dagger$ and $2 \difz = \npz
\npz^\dagger$ imply that
\begin{align} \label{eq:D1}
  2 \delta \dif = \npz \npd^\dagger + \npd \npz^\dagger \ .
\end{align}
Note that one should not expect to solve this equation for $\npd$
uniquely given a specific $\delta \dif$; many noise coefficient
matrices $\npf$ map into the same diffusion matrix $\dif$.

The relation above can now be used to express the perturbation
parameter $\hh = \npd \npz^{-1}$ in terms of $\difz$ and $\delta
\dif$, but not uniquely, which is not a problem. Multiplying
\Eqref{eq:D1} by $\difz^{-1}$ from the left and right, one easily
finds
\begin{align}
  \difz^{-1} \delta \dif \difz^{-1} = \difz^{-1} \hh + \hh^\dagger
  \difz^{-1}\ .
\end{align}
The left hand side of this equation is equal to $-\delta \dif^{-1}$
(verified easily by varying the identity $1 = \dif \dif^{-1}$) which
is determined by the physical description of the
perturbation. Meanwhile, the right hand side of the equation is twice
the symmetric part of $\difz^{-1} \hh$. In other words, it is only the
symmetric part of $\difz^{-1} \hh$ that is fixed by the physical form
of the perturbation, and the antisymmetric part is left
undetermined. It therefore behooves one to choose $\difz^{-1} \hh$ to
be purely symmetric, whence one obtains $-\delta \dif^{-1} = 2
\difz^{-1} \hh$, or
\begin{align}
  \hh = -\tfrac 12 \difz \delta \dif^{-1}
  = -\tfrac 12 \dif \delta \dif^{-1} \ ,
\end{align}
where the second equality is valid due to the overarching first order
approximation of linear response.

\section{Asymptotic values of the susceptibility} \label{app:asy}
In this appendix, we sketch how the asymptotic values for the
susceptibilities in Figures \ref{fig:U} and \ref{fig:U2} were
obtained.

One can derive a host of relations valid in the stationary regime by
requiring that time derivatives of state observables vanish
\cite{falvir}. For the present discussion, the relevant observable is
the tensor $\xx_i \xx_j$, \ie.,
\begin{align}
  0 = \frac{\dd}{\dd t} \mean{\xx_i \xx_j} = \mean{\Lgen \xx_i \xx_j} \ ,
\end{align}
with the backward generator $\Lgen = \dri_i \del_i + \dif_{ij} \del_i
\del_j$. (One should keep in mind that the average is in the
stationary regime, although we leave it unlabeled.) When evaluated
explicitly, with $\dri = \mob \for$, one finds the relation
\begin{align}
  0 = \mob \mean{\for \xx^\dagger} + \mean{\xx \for^\dagger} \mob
  + 2 \dif \ . \label{eq:virial}
\end{align}
This matrix relation entails a set of equations for the independent
components of the tensor $\mean{\xx_i \for_j}$, of which there are 3
in our case with 2 degrees of freedom. 

We note that for our system, $U(\xx) = -(1/2) \Tr (\for \xx)$. Thus,
multiplying \Eqref{eq:virial} from the left by $\mob^{-1}$ and taking
the trace, we find the stationary average of the potential energy to
be
\begin{align}
  \mean{U(\xx)} = -\frac 12 \Tr \mean{\for \xx} 
  = \frac 12 \Tr(\mob^{-1} \dif) \ .
\end{align}
Hence, the stationary (asymptotic) value for the susceptibility is
found as
\begin{align}
  \frac {\del \mean{U}} {\del \Tem}
  = \frac 12 \Tr \! \left( \mob^{-1} \frac{\del \dif}{\del \Tem}\right)
  = \frac 12 \ , \label{eq:sus.asy}
\end{align}
which was evaluated using \Eqref{eq:dif.2} for the diffusion
matrix. This asymptotic value for $\sus$ was indicated in Figure \ref{fig:U}.

The susceptibility of the energy of the second particle $U_2 (\xx_2) =
-(1/2) \for_2 \xx_2$ can be extracted similarly from
\Eqref{eq:virial}, with the exception that one has to go through the
tedium of actually solving for the component $\mean{\for_2 \xx_2}$. We
quote only the result of this straightforward exercise:
\begin{align}
  \frac {\del \mean{U_2}} {\del \Tem} =& -\frac 12 \frac{\del}{\del
    \Tem} \mean{\for_2 \xx_2} = \frac {\varepsilon^2}
        {2\left(1+\frac{\kappa_{11}}{\kappa_{22}}\right)}
        \ . \label{eq:sus2.asy}
\end{align}
This was evaluated as $0.01899$ for the actual experimental values of
$\kappa_{11} = \SI{3.3745}{\pico\newton/\micro\meter} $, 
$\kappa_{22} = \SI{3.3285}{\pico\newton/\micro\meter} $,
and $\varepsilon = 0.2766$, and indicated in Figure
\ref{fig:U2} as the asymptotic value of the susceptibility $\sus$.

\bibliographystyle{spmpsci}


\end{document}